\documentclass[showpacs,notitlepage,superscriptaddress,prd,aps,twocolumn,nofootinbib]{revtex4}
\usepackage{graphicx}\usepackage{amsmath,amssymb}\usepackage{slashed}
\usepackage{epstopdf}
\usepackage{adjustbox}

\newcommand{\be}{\begin{equation}}
\newcommand{\ee}{\end{equation}}
\newcommand{\bea}{\begin{eqnarray}}
\newcommand{\eea}{\end{eqnarray}}
\newcommand{\ii}{\mathrm{i}}

\newcommand{\dd}{\mathrm{d}}

\newcommand{\tM}{\widetilde{\mathcal{M}}}
\newcommand{\tD}{\widetilde{D}}

\begin{document}
\title{Parity anomaly in four dimensions}
\affiliation{CMCC-Universidade Federal do ABC, Santo Andr\'e, S.P., Brazil}
\affiliation{Department of Physics, Tomsk State University, 634050 Tomsk, Russia}

\author{M. Kurkov}
\email{max.kurkov@gmail.com}
\affiliation{CMCC-Universidade Federal do ABC, Santo Andr\'e, S.P., Brazil}

\author{D. Vassilevich}
\email{dvassil@gmail.com}
\affiliation{CMCC-Universidade Federal do ABC, Santo Andr\'e, S.P., Brazil}
\affiliation{Department of Physics, Tomsk State University, 634050 Tomsk, Russia}
\begin{abstract}
In an analogy to the odd-dimensional case we define the parity anomaly as the part of the one-loop effective action for fermions associated with spectral asymmetry of the Dirac operator. This quantity is computed directly on four-dimensional manifolds with boundary and related to the Chern-Simons current on the boundary. Despite a quite unusual Chern-Simons level obtained, the action is gauge invariant and passes all consistency checks.
\end{abstract}
\maketitle

\section{Introduction}
The anomalous parity symmetry breaking in quantum theories of fermions in odd-dimensional spaces (the parity anomaly) was discovered in mid 1980's \cite{Niemi:1983rq,Redlich:1983kn,AlvarezGaume:1984nf}. In three dimensions, in particular, the Chern-Simons action with the level $\pm \tfrac 12$ is induced by this mechanism by one generation of Dirac fermions. As a good source on Chern-Simons theories the reader may consult Ref.\ \cite{Dunne:1998qy}.

This topic received a renewed interest recently in a relation to Topological Insulators and other topological materials (see, \cite{Tkachov:2016}, for example). In this context, the Chern-Simons action describes Hall conductivity of the surface modes. Physical implications of the parity anomaly for Topological Insulators have been discussed in a number of papers \cite{Mulligan:2013he,Zirnstein:2013tba,Konig:2014ema}. To understand this situation it is important to perform quantum field theory computations directly on a four-dimensional manifold with boundaries. It is somewhat surprising that no one has done this before.

The purpose of this paper is to compute the parity anomaly on a Euclidean manifold for Dirac fermions interacting with a $U(1)$ gauge field. We impose the local bag boundary conditions on the spinors that ensure that the current through the boundary vanishes, so that the fermions are confined in the bulk. The parity anomaly is understood as the part of the effective action related to the spectral asymmetry of the Dirac operator. We use the spectral methods and heat kernel expansion \cite{Kirsten:2001wz,Vassilevich:2003xt,Gilkey:2004dm} that proved their efficiency in three dimensions.

This paper is organized as follows. In the next Section we introduce the main conventions and define the boundary conditions. In Sec.\ \ref{sec:par} we use the the $\zeta$ function regularization to compute the variation of the parity odd part of the effective action with respect to the external electromagnetic field. This variation has the form a boundary Chern-Simons current. If the electromagnetic field belongs to a trivial $U(1)$ bundle, the variation can be integrated to a boundary Chern-Simons action with an unusual level $k=\pm \frac 14$. The computation of the parity odd effective action for topologically non-trivial stationary configurations are contained in Sec.\ \ref{sec:stat}. The subsequent Section is dedicated to various consistency checks, that includes the Laughlin argument and relations to the surface mode counting. The results a briefly discussed in Sec.\ \ref{sec:Con}. Useful formulas on the heat kernel expansion are collected in Appendix \ref{App:HK}.

\section{The set-up}\label{sec:set}
We consider the Dirac operator 
\be
\slashed{D} = \ii \gamma^{\mu}\left(\nabla_{\mu} + \ii A_{\mu}\right) \label{Dop}
\ee
with an abelian gauge field $A_\mu$ on a four-dimensional Euclidean manifold $\mathcal{M}$ with a boundary $\partial\mathcal{M}=\bigcup_\alpha \partial\mathcal{M}_\alpha$ consisting of several connected components $\partial\mathcal{M}_\alpha$ numbered by the index $\alpha$. The gamma matrices satisfy the relation
\be
\{\gamma^{\mu},\gamma^{\nu}\} = g^{\mu\nu}1_4, \label{gam}
\ee
where $1_4$ is a unit matrix in the spinor indices, $g^{\mu\nu}$ is the Riemannian metric and $\nabla$ is a covariant derivative with the spin-connection compatible with $g$.

We assume that the manifold $\mathcal{M}$ is flat, though the boundaries may be curved. Near the boundaries the natural coordinate system implies a non-constant metric and a spin connection term in the Dirac operator. 

The chirality matrix $\gamma_5$ is defined through the Levi-Civita tensor as
\begin{equation}
\gamma^5 = \tfrac{1}{4!} \epsilon^{\mu\nu\rho\sigma}\gamma_\mu \gamma_\nu \gamma_\rho \gamma_\sigma \,.
\end{equation}

Since there are boundaries, we have to impose some boundary conditions on the spinor field $\psi$. Let $n$ be the inward pointing unit normal to the boundary. Define the projector
\begin{equation}
\Pi_-:=\tfrac 12 \bigl( 1 - \ii \varepsilon_\alpha \gamma^5\gamma^n \bigr)\,. \label{Pim}
\end{equation}
The sign factor $\varepsilon_\alpha=\pm 1$ may vary from one component of the boundary to another being constant on each of the components. We take the boundary conditions in the form
\begin{equation}
\Pi_-\psi \vert_{\partial\mathcal{M}}=0\,,\label{bag}
\end{equation}
which is nothing else then the (Euclidean version of) bag boundary conditions proposed first in \cite{Chodos:1974je} and then rediscovered in the mathematical context in Ref.\ \cite{Branson:1992}. Later on, to compute the heat trace asymptotics of $\slashed{D}^2$, we shall also need boundary conditions for the second half of the spinor components. These conditions follow from the consistency of (\ref{bag}) and read
\begin{equation}
\Pi_-\slashed{D}\psi\vert_{\partial\mathcal{M}}=0\,.\label{bag2}
\end{equation}
 By commuting $\Pi_-$ through the Dirac operator and using again the Dirichlet condition (\ref{bag}) we may rewrite (\ref{bag2}) in the form of a Robin condition
\begin{equation}
(\nabla_n +iA_n+S)\Pi_+\psi\vert_{\partial\mathcal{M}}=0\,,\label{bag22}
\end{equation}
where
\begin{equation}
\Pi_+=\tfrac 12 \bigl( 1 + \ii \varepsilon_\alpha \gamma^5\gamma^n \bigr)\,,\qquad
S=-\tfrac 12 \Pi_+ K \label{Pip}
\end{equation}
with $K$ being the trace of extrinsic curvature of the boundary.

These boundary conditions yield a vanishing current through the boundary, $j^n\vert_{\partial\mathcal{M}} = \bar\psi \gamma^n \psi \vert_{\partial\mathcal{M}}=0$. Besides, with these boundary conditions $\slashed{D}$ is symmetric and the boundary value problem is strongly elliptic.

\section{Parity anomaly}\label{sec:par}
Let us consider the one-loop effective action for fermion in the presence of an external electromagentic vector potential. At the beginning, when the treatment is independent of the dimension and of the presence or absence of boundaries, we shall follow the monograph \cite{Fursaev:2011zz} (which in turn used \cite{AlvarezGaume:1984nf,Deser:1997nv}). Only the main steps will be repeated. The zeta function of $\slashed{D}$ is defined as usual through summation over the eigenvalues $\lambda$,
\begin{equation}
\zeta (s,\slashed{D})=\sum_{\lambda>0} \lambda^{-s} + e^{-\ii \pi s}
\sum_{\lambda<0} (-\lambda)^{-s} \label{zetaD}
\end{equation}
Here $s$ is a complex parameter. The sums above are convergent for $\Re s$ large enough. $\zeta(s,\slashed{D})$ may be continued as a meromorphic function to the whole complex plane. 

We can separate in two parts that are even and odd with respect to the reflection $\slashed{D}\to -\slashed{D}$, respectively:
\begin{eqnarray}
&&\zeta(s,\slashed{D})=\zeta(s,\slashed{D})_{\rm even} +\zeta(s,\slashed{D})_{\rm odd}\,,\nonumber\\
&&\zeta(s,\slashed{D})_{\rm even} =\tfrac 12 \bigl( \zeta(s,\slashed{D})+\zeta(s,-\slashed{D})\bigr)\,,\label{zeven}\\
&&\zeta(s,\slashed{D})_{\rm odd} =\tfrac 12 \bigl( \zeta(s,\slashed{D})-\zeta(s,-\slashed{D})\bigr)\,.\label{zodd}
\end{eqnarray}
The even part will not be discussed in this paper.
The odd term can be rewritten through the eta function
\begin{equation}
\zeta(s,\slashed{D})_{\rm odd}=\tfrac 12 \bigl(1-e^{-\ii \pi s}\bigr)\eta(s,\slashed{D}) \,,\label{oddeta}
\end{equation}  
which is, by definition,
\begin{equation}
\eta(s,\slashed{D}):=\sum_{\lambda>0} \lambda^{-s} -
\sum_{\lambda<0} (-\lambda)^{-s}\,.\label{eta}
\end{equation}

Let us recall the definition of zeta-regularized determinant of $\slashed{D}$ and the zeta regularized one-loop effective action: 
\begin{equation}
W_s=-\ln\det (\slashed{D})_s=\mu^s \Gamma(s)\zeta(s,\slashed{D}). \label{zreg}
\end{equation}
$\mu$ is a parameter of the dimension of the mass which makes the whole expression above dimensionless.
This effective action can be separated in even and odd parts, $W^{\rm even}_s$ and $W^{\rm odd}_s$ corresponding to the even and odd parts of the zeta function.  
In the limit $s\to 0$, which corresponds to removal of the regularization, the part $W^{\rm odd}_s$ remains finite as the root of the prefactor in (\ref{oddeta}) cancels the pole in the $\Gamma$-function. We have therefore
\begin{equation}
W^{\rm odd}\equiv W^{\rm odd}_{s=0}=\frac{\ii \pi}2 \eta(0,\slashed{D}).\label{Wodd}
\end{equation}

Let us make use of the following integral representation for the eta function
\be
\eta(s,\slashed{D}) = \frac{2}{\Gamma\bigl(\tfrac{s+1}{2}\bigr)}\int^{\infty}_0 d\tau \, \tau^{s} \,\mathrm{Tr}\,\left(\slashed{D}e^{-\tau^2\slashed{D}^2}\right)
\ee
through the trace of the heat operator $e^{-\tau^2\slashed{D}^2}$. Under small variations $A_\mu\to A_\mu + \delta A_\mu$ the eta functions varies as 
\be
\delta\eta(s,\slashed{D}) = \frac{2}{\Gamma\bigl(\tfrac {s+1}2\bigr)}\int^{\infty}_0 d\tau \, \tau^{s} \, 
\frac{\dd}{\dd \tau }\mathrm{Tr}\,\left((\delta \slashed{D}) \tau e^{-\tau^2\slashed{D}^2}\right). \nonumber
\ee
In this formula, one may set $s=0$. Since the heat trace weighted with a zeroth order operator $\delta \slashed{D}=-\gamma^\mu \delta A_\mu$ vanishes sufficiently fast at $\tau\to \infty$, we may write
\begin{equation}
\delta \eta(0,\slashed{D})=-\frac 2{\pi^{1/2}} \lim_{t\to +0} \mathrm{Tr}\,\left((\delta \slashed{D})t^{1/2}e^{-t\slashed{D}^2}\right).\label{deta0}
\end{equation}
(Here $t=\tau^2$). 

For any smooth matrix-valued function $Q$ (more scientifically - for an endomorphism of the spin bundle) there is a full asymptotic expansion at $t\to +0$
\begin{equation}
\mathrm{Tr}\,\left(Q e^{-t\slashed{D}^2}\right)\simeq \sum_{k=0}^\infty t^{\frac{k-4}2}a_k(Q,\slashed{D}^2).
\label{asymp}
\end{equation}
Clearly, only the coefficients $a_k(Q,\slashed{D}^2)$ with $k\leq 3$ may contribute to (\ref{deta0}). Generic expressions for these coefficients are given in Appendix \ref{App:HK}. By substituting in these expressions the particular values of the invariants corresponding to our problem and computing the traces, one obtains that for $k<3$ all coefficients $a_k$ vanish, so that (\ref{deta0}) remains finite in the limit $t\to 0$. In $a_3$, see (\ref{a3}), just two terms,
$48E\chi+48\chi E$, contribute, so that
\begin{equation}
a_3(Q,\slashed{D}^2)=\frac 1{8\pi^{3/2}} \int_{\partial\mathcal{M}} d^3x\sqrt{h} \varepsilon_\alpha\epsilon^{nabc}
(\delta A_a)\partial_bA_c. \label{a3final}
\end{equation}
We recall that $Q=\delta\slashed{D}=-\gamma^\mu\delta A_\mu$, the index $n$ corresponds to the inward pointing unit normal, $a,b,c$ denote directions tangential to the boundary. Other notations are explained in Appendix \ref{App:HK}.

Next, we substitute (\ref{a3final}) in (\ref{deta0}) and (\ref{Wodd}) to obtain
\begin{equation}
\delta W^{\rm odd}=-\frac {\ii}{8\pi} \int_{\partial\mathcal{M}} d^3x\sqrt{h} \varepsilon_\alpha\epsilon^{nabc}
(\delta A_a)\partial_bA_c \,.\label{varWodd}
\end{equation}
This is the main result of this Section.

Before proceeding further, let us stress that the spectral methods used here are applicable whenever the variation $\delta A$ and the field strength $F_{\mu\nu}$ of $A_\mu$ are smooth everywhere on $\mathcal{M}$. The field $A_\mu$ itself has to be smooth up to a gauge transformation. This excludes, for example, monopoles in $\mathcal{M}$. However, the configurations corresponding to monopoles outside $\mathcal{M}$ are allowed. If, e.g., $\mathcal{M}$ is a finite thickness spherical shell times the Euclidean time direction compactified on $S^1$, one can place a magnetic charge in the center of the shell.

Since smooth localized variations are always allowed, one can derive the boundary current corresponding to $W^{\rm odd}$
\begin{equation}
j^a_{\rm odd}=-\frac {\ii}{8\pi}\varepsilon_\alpha\epsilon^{nabc}
\partial_bA_c \,.\label{current}
\end{equation}
This current corresponds to an antisymmetric, Hall type, conductivity tensor.

Consider next the case when the gauge potential $A_\mu$ itself is smooth (a trivial $U(1)$ bundle). Then, this configuration can be connected by a smooth homotopy to the one with $A_\mu =0$. One can impose the natural condition $W^{\rm odd}(A=0)=0$
and resolve the variational equation (\ref{varWodd}) to obtain
\begin{equation}
W^{\rm odd}(A)=-\frac {\ii}{16\pi} \int_{\partial\mathcal{M}} d^3x\sqrt{h} \varepsilon_\alpha\epsilon^{nabc}
A_a\partial_bA_c \,.\label{Woddfin}
\end{equation}
Thus, each boundary component $\partial\mathcal{M}_\alpha$ carries the Chern-Simons action\footnote{Since $\epsilon^{nabc}$ is the Levi-Civita tensor rather than the Levi-Civita symbol, the action (\ref{Woddfin}) does not depend on the metric.} with a quite unusual level $k=\tfrac 14 \varepsilon_\alpha$.

We like to stress that the expression (\ref{Woddfin}) is \emph{not} valid for topologically non-trivial configurations of the gauge field in general. The technical reason is that one cannot integrate by parts in the Chern-Simons action if the gauge potential is not continuous. See the next Section for a discussion of topologically non-trivial configurations.

Note, that by definition the zero modes of Dirac operator do not contribute to the spectral functions (\ref{zetaD}) and (\ref{eta}). Thus zero modes also have to be excluded from the heat kernel. This, however, does not affect our result (\ref{Woddfin}) since the exclusion of any finite number of modes does not change the coefficients $a_k$, $k\leq 3$, corresponding to negative powers of the proper time $t$ in the expansion (\ref{asymp}).

We conclude this section with a short discussion of what kind of classical symmetry corresponds to the parity anomaly in four dimensions. The transformation $\psi\to \gamma^5\psi$ together with the inversion $\varepsilon_\alpha \to - \varepsilon_\alpha$ on all components of the boundary maps solutions of the classical equation $\slashed{D}\psi =0$ to other solutions. However, this transformation inverts the non-zero spectrum of Dirac operator. Thus, this classical symmetry cannot be carried on to the quantum theory.

\section{Stationary configurations}\label{sec:stat}
Here we compute exactly the parity anomaly for static gauge fields on a kind of finite-temperature manifolds (the ones having an $S^1$ as a factor). There is a large literature with similar calculations in three dimensions, see e.g. \cite{Niemi:1986kh,Deser:1997nv,Fosco:2017vxl}. We got our inspiration from \cite{Deser:1997gp}.

Let us consider $\mathcal{M}=\tM \times S^1$ with all gauge fields being constant with respect to the Euclidean time coordinate $x^4$ along $S^1$ which changes from $0$ to $2\pi$. We assume that $S^1$ has the unit radius and that $a:=A_4$ does not depend on the coordinates of $\tM$.

Let us split the Dirac operator as
\begin{equation}
\slashed{D}=\ii \gamma^4 (\partial_4+\ii a) +\tD \label{tD}
\end{equation}
and denote by $\psi (\mu)$ the eigenmodes of $\tD$,
\begin{equation}
\tD \psi (\mu)=\mu \psi (\mu) \,.
\end{equation}
The matrix $\gamma^4$ commutes with the boundary projectors, $[\gamma^4,\Pi_\pm]=0$, and anticommutes with $\tD$, $\gamma^4\tD=-\tD \gamma^4$. Therefore, all eigenmodes with $\mu\neq 0$ may be combined in pairs $\psi(\mu),\ \psi(-\mu)$ such that $\gamma^4\psi(\mu)=\psi (-\mu)$. When acting on such a pair, the operator $\slashed{D}$ has the form
\begin{equation}
\slashed{D}=\left( \begin{array}{cc} \mu & -(\omega +a) \\ -(\omega+a) & -\mu
\end{array} \right)\,, \label{sDD}
\end{equation}
where $\omega$ denotes the eigenfrequencies in the $x^4$ direction, $\partial_4\psi(\pm\mu)=\ii\omega\psi(\pm\mu)$.
The eigenvalues of $\slashed{D}$ read
\begin{equation}
\lambda =\pm \sqrt{\mu^2+(\omega +a)^2} \,.\label{squareroot}
\end{equation}
We see, that the eigenvalues always appear in pairs of opposite sign. Therefore, there is no contribution to the $\eta$-function (\ref{eta}) and to $W^{\rm odd}$ from $\psi (\mu)$ with $\mu\neq 0$.

The situation is different with the zero modes of $\tD$. The matrix $\gamma^4$ maps the zero eigenspace onto itself. Therefore, all eigenmodes of $\tD$ can be separated into positive and negative modes of $\gamma^4$, $\psi (0)_+$ and $\psi(0)_-$, so that $\gamma^4\psi (0)_\pm =\pm \psi(0)_\pm$. 
\begin{equation}
\slashed{D}\psi (0)_\pm = \mp (\omega + a) \psi (0)_\pm \label{sDp0}
\end{equation}
The numbers of $\psi(0)_+$ and $\psi(0)_-$ modes ($n_+$ and $n_-$, respectively) are in general different. The difference is given by the $\gamma^4$-index of $\tD$,
\begin{equation}
n_+-n_-={\rm Ind}\, (\tD) \label{Ind1}
\end{equation}
that may be written as
\begin{equation}
{\rm Ind}\, (\tD)=\lim_{t\to 0} {\rm Tr}\, \left( \gamma^4 e^{-t\tD^2} \right)
\label{Ind2}
\end{equation}
Note, that in these formulas the operator $\tD$ has to be considered as an operator on $\tM$ rather than on $\mathcal{M}$. The index can be computed with the help of the heat kernel expansion, see Appendix \ref{App:HK}.
\begin{equation}
{\rm Ind}\, (\tD)=a_3(\gamma^4,\tD^2)\,,
\end{equation}
where
\begin{equation}
a_3(\gamma^4,\tD^2)=-\frac {1}{4\pi} \int_{\partial\tM} d^2x\sqrt{\tilde h} \varepsilon_\alpha\epsilon^{n4bc}\partial_bA_c \,.\label{a3tD}
\end{equation}

Let us impose the periodic condition on $\psi$, $\psi\vert_{x^4}=\psi\vert_{x^4+2\pi}$. Then the frequencies $\omega$ are integers. The effective actions depends on the non-integer part of $a$, $\bar a=a-\left\lfloor a \right\rfloor$. Thus the contribution of $\psi(0)_-$ modes to the $\eta$-function (\ref{eta}) reads
\begin{eqnarray}
\eta_-(s)& = &\sum_{\omega\in \mathbb{N}_0} (\omega + \bar a)^{-s} -
\sum_{\omega\in -\mathbb{N}} (-\omega - \bar a)^{-s} \nonumber\\
&=&\zeta_R(s,\bar a) -\zeta_R(s,1-\bar a) \,.\label{etazeta}
\end{eqnarray}
Here $\zeta_R(s,\bar a)$ is the generalized Riemann (Hurwitz) zeta function. At $s=0$
\begin{equation}
\zeta_R(0,\bar a)=\tfrac 12 -\bar a \label{zetaR}
\end{equation}

The contribution of $\psi(0)_+$ to the $\eta$-function is just opposite to that of $\psi(0)_-$, $\eta(0)_+=-\eta(0)_-$. Thus the whole $\eta(0)$ reads
\begin{equation}
\eta(0)=-{\rm Ind}\,(\tD)\, \eta(0)_-=(2\bar a-1){\rm Ind}\,(\tD)\,.
\end{equation}
Collecting everything together, we arrive at the following expression for the effective action
\begin{equation}
W^{\rm odd}=\frac{\ii (1-2\bar a)}8 \int_{\partial\tM} d^2x\sqrt{\tilde h} \varepsilon_\alpha\epsilon^{n4bc}\partial_bA_c \,.\label{Wstat}
\end{equation}

Since $S^1$ has a non-trivial fundamental group $\pi_1$, there are large gauge transformations $\psi\to e^{-\ii x^4 l}\psi$, $a\to a+l$ with $l\in \mathbb{Z}$. These transfromations leave $\bar a$ invariant and thus do not change the action (\ref{Wstat}).

The index of $\tD$ may be expressed as
\begin{equation}
{\rm Ind}\,(\tD)=\sum_\alpha \varepsilon_\alpha M_\alpha [A] ,\label{indflux}
\end{equation}
where $M_\alpha [A]$ is the magnetic flux though the boundary component $\tM_\alpha$. Magnetic charges are not allowed inside $\mathcal{M}$. Therefore, if all $\varepsilon_\alpha$ are equal, the total magnetic flux is zero, and ${\rm Ind}\,(\tD)=0$ as well. If one flips the sign of one of the $\varepsilon_\alpha$, the index changes by twice the flux though corresponding component of the boundary, i.e., by an even number. We conclude, that always
\begin{equation}
{\rm Ind}\,(\tD)=2N, \qquad N\in \mathbb{Z}\,.\label{Ind2N}
\end{equation}
Let us replace $\bar a$ with $\bar a+1$ in (\ref{Wstat}). Due to (\ref{Ind2N}) the effective action is shifted as
\begin{equation}
W^{\rm odd}\to W^{\rm odd} + 2\pi N \ii \,.\label{WtoW}
\end{equation}
Thus, the partition function remains unchanged under this transformation, and one can simply remove the bar over $a$ in Eq.\ (\ref{Wstat}).

An important consistency check is that the variation of (\ref{Wstat}) with respect to $a$ coincides with the variational equation (\ref{varWodd}). 

Note, that naive substitution of the gauge field configuration considered in this Section into the action (\ref{Woddfin}) (that has been derived to topologically trivial configurations) would give just one half of the correct result (\ref{Wstat}). At the same time, (\ref{Wstat}) vanishes on topologically 
trivial gauge configurations. Thus, one may in principle write an interpolating formula for the effective action valid for both trivial and stationary non-trivial $U(1)$ bundles. Such a formula would be, however, of a limited use since on cannot guarantee its' validity for gauge configurations that are non-stationary and topologically non-trivial at the same time.

It is interesting to consider the case of anti-periodic conditions for $\psi$ on $S^1$ corresponding to finite temperature fermions. In this case, $\omega\in\mathbb{N}+\tfrac 12$. As one can easily see, the formulas in this Section, including the action (\ref{Wstat}), remain valid after the replacement $\bar a\to \bar a+\tfrac 12$.

\section{Smooth gauge potentials}\label{sec:smo}
\subsection{Gauge invariance and Laughlin argument}
\begin{figure}[ht]
\begin{center}
\adjustbox{trim={.22\width} {.30\height} {0.2\width} {.30\height},clip}
{\includegraphics[width=12cm]{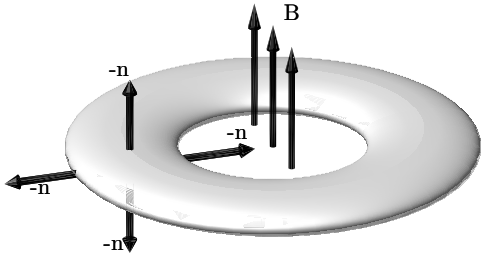}}
\caption{The geometry for testing the Laughlin argument. $B$ is the magnetic field piercing the whole of the torus. $-n$ is the outward pointing normal to the surface torus at different points. We remind, that $n$ is the inward pointing normal vector in our notations.
\label{fig}}
\end{center}
\end{figure}

In this work we have used the $\zeta$-function regularization that manifestly preserves the gauge symmetries. However, we find it instructive to see how the celebrated Laughlin argument works in our set-up. This argument uses insertion of a magnetic flux in a geometry containing a matter sample and restricts possible values of the Hall conductivity. Originally, this procedure was formulated for a circular ribbon \cite{Laughlin:1981jd} and for an annular film \cite{Halperin:1981ug}, both being purely 2-dimensional samples. Since we need a 3-dimensional bulk with a 2-dimensional surface, we shall use the version suggested in \cite{Konig:2014ema} (see also \cite{Vafek:2011zh}). The geometry consists of a squashed doughnut, see Fig.\ \ref{fig}, with a time-dependent magnetic flux piercing the hole.
The consistency of this picture requires \cite{Konig:2014ema} the sum of the Hall conductivities corresponding to the currents (\ref{current}) on the top and on the bottom of the doughnut to be an integer multiplier of $e^2/h$. (According to our conventions, the elementary charge $e=1$ and $\hbar=h/(2\pi)=1$). This requirement is trivially satisfied in our case. Indeed, we have a single boundary component and a single value of $\varepsilon$ everywhere. The normal vectors are opposite on the top and on the bottom, so that the sum of the conductivities is zero.

To avoid confusions, we stress that to ensure validity Eq. (\ref{Woddfin}) the gauge potential needs to be smooth just on $\mathcal{M}$, but may have discontinuities in the ambient space.

\subsection{Relations to surface states}
We have obtained, see Eq.\ (\ref{Woddfin}), the Chern-Simons action with the level $k$,
$|k|=\tfrac 14$, on each of the components of the boundary if the gauge field belongs to a trivial $U(1)$ bundle. This value of $|k|$ may seem too small as the 3D calculations give $|k|=\tfrac 12$, or too large since there are no surface modes of massless bulk fermions for some of the geometries (like the half-space, e.g.). Here we show that our result is indeed consistent with a kind of the Kaluza-Klein 3D limit of the 4D manifolds.

To simplify the analysis, we stretch a bit our results to the non-compact setting.
We consider the geometry of a slab with infinite parallel boundaries at $x^4=0$ and $x^4=l$. A Kaluza-Klein type limit $l\to 0$ will be considered. 

The spectrum of free Dirac operator in this space can be easily found. 
For the boundary conditions with opposite values of $\varepsilon$ on two boundaries
\begin{equation}
\lambda^2(k_a,p)=k_a^2 + \frac{\pi^2 p^2}{l^2} \label{spec1}
\end{equation}
where $p$ is an integer, and $k_a$ is the 3-momentum along the boundaries. From the 3D point of view, this spectrum corresponds to a tower of states with the masses $m_p^2=\pi^2 p^2 /l^2$. In the limit $l\to 0$ the state with $p=0$ remains, that correspond to a massless fermion in 3D. Such a field in 3D has to generate a Chern-Simons term with $|k|=\tfrac 12$, which is precisely what the action (\ref{Woddfin}) gives in the limit $l\to 0$ for opposite values of $\varepsilon$. (Note, that the normal vectors on two components of the boundary are also opposite).

For equal values of $\varepsilon$ at $x^4=0$ and $x^4=l$, the spectrum reads
\begin{equation}
\lambda^2(k_a,p)=k_a^2 + \frac{\pi^2 \big( p+ \tfrac 12 \big)^2}{l^2} \,.\label{spec2}
\end{equation}
We see, that in the limit $l\to 0$ all states become infinitely massive. This is consistent with the fact that the Chern-Simons actions in (\ref{Woddfin}) on two components of the boundary cancel against each other in this limit.

Let us comment briefly on the case of the half-space. It is true that the massless Dirac operator with our boundary conditions does not have eigenmodes that decay exponentially away from the boundary (i.e., the edge states). However, this does not yet mean that the effective action does not have any boundary contributions. At any rate, it would be interesting to perform direct computations of the parity anomaly on the half-space, which is the simplest non-compact manifold with a boundary. 

Non-uniqueness of the induced Chern-Simons action in the Kaluza-Klein limit (that appears to depend on the boundary conditions) may have something to do with the ambiguity in the parity anomaly for lattice fermions discovered in \cite{Coste:1989wf}.

\section{Conclusions}\label{sec:Con}
Let us briefly summarize the main findings of this paper. We considered one generation of massless Dirac fermions on a Euclidean manifold with boundaries subject to bag boundary conditions. We studied the parity anomaly for these fermions, which is the part of the effective action associated to the spectral asymmetry of the Dirac operator. For a generic abelian gauge field we computed the variation of the parity odd effective action (\ref{varWodd}) and the induced Chern-Simons current (\ref{current}). The action itself was obtained in two particular cases. For stationary topologically nontrivial gauge field configurations the action was computed in Sec.\ \ref{sec:stat}. For trivial $U(1)$ bundles, the variation was integrated to the Chern-Simons action (\ref{Woddfin}) with an unusual level $k=\pm \tfrac 14$. This result passed several consistency checks in Sec.\ref{sec:smo}. 

An important message of the present work is that due to the spectral asymmetry of Dirac operator on 4D manifolds with boundaries one cannot use $\slashed{D}^2$ alone to construct the full effective action. A similar conclusion (though in a different set-up) has been obtained in the recent work \cite{Kalinichenko:2017nbz}.

Existing computations of the parity anomaly from four dimensional theories deal with domain walls rather than with boundaries, see \cite{Mulligan:2013he}. The spectral asymptotics of the Dirac and Laplace operators for the configurations with domains walls and with boundaries are quite different, so that there cannot be any direct comparison with our results.

A natural question is whether this parity anomaly is measurable. The answer is probably positive, though the Wick rotation to the Minkowski signature on manifolds with boundaries is not a simple problem. Another hurdle is that in all real materials the fermionic excitations excitations have some chemical potential, probably a mass gap, and definitely they experience some scattering on impurities. This just stresses the importance of Minkowski signature computations with all the effects listed above turned on, even for the simplest geometries.

It is both interesting and important to compute the parity anomaly for other types of the boundary conditions, that may include the chiral bag conditions (that have a chiral phase on the boundary) or even the non-local Atiyah-Patodi-Singer conditions. Unfortunately, the relevant heat kernel coefficient $a_3(Q,\slashed{D}^2)$ is not known yet for these two types of boundary conditions.

\acknowledgments 
We are grateful to Ignat Fialkovsky for helpful remarks. This work was supported by the grants 2015/05120-0 and 2016/03319-6 of the S\~ao Paulo Research Foundation (FAPESP),  by the grants 401180/2014-0 and 303807/2016-4 of CNPq, and by the Tomsk State University Competitiveness Improvement Program.


\appendix
\section{Heat kernel expansion}\label{App:HK}
Boundary conditions (\ref{bag2}), (\ref{bag22}) belong to the so called mixed type. Spectral asymptotics of these boundary conditions were thoroughly studied in \cite{mixedBC}.
The relevant term in the heat kernel expansion will be taken from \cite{Marachevsky:2003zb} after a suitable adjustment of the notations.  One can write $\slashed{D}^2$ in the canonical form of Laplace type operators:
\begin{equation}
\slashed{D}^2=-(\hat\nabla^2 +E)\,,\label{D2}
\end{equation}
where the covariant derivative $\hat\nabla =\partial +\omega$ and
\begin{eqnarray}
&&\omega_\mu = \ii A_\mu +\tfrac 18 [\gamma_\nu,\gamma_\rho]\sigma_\mu^{[\nu,\rho]}\,,\\
&&E=\tfrac {\ii}4 [\gamma^\mu,\gamma^\nu]F_{\mu\nu}, \qquad F_{\mu\nu}=\partial_\mu A_\nu - \partial_\nu A_\mu\,,
\end{eqnarray}
$\sigma_\mu$ is the spin connection. 

Let us now list the coefficients appearing in the asymptotic expansion (\ref{asymp})
\begin{eqnarray}
&&a_0(Q,\slashed{D}^2)=\frac 1{(4\pi)^{m/2}} \int_{\mathcal{M}} d^mx\sqrt{g} {\rm tr}\, Q \nonumber\\
&&a_1(Q,\slashed{D}^2)=\frac 1{4\cdot (4\pi)^{(m-1)/2}} \int_{\partial\mathcal{M}} d^{m-1}x\sqrt{h}{\rm tr}\,(\chi Q)\nonumber \\
&&a_2(Q,\slashed{D}^2)=\frac 1{6\cdot (4\pi)^{m/2}} \left\{ \int_{\mathcal{M}} d^mx\sqrt{g} {\rm tr}\, Q E + \right.\nonumber \\
&& \qquad + \left. \int_{\partial\mathcal{M}} d^{m-1}x\sqrt{h}{\rm tr}\, (2QK + 12 QS+3Q_{;n}) \right\}\nonumber \\
&&a_3(Q,\slashed{D}^2)=\frac 1{384\cdot (4\pi)^{(m-1)/2}} \int_{\partial\mathcal{M}} d^{m-1}x\sqrt{h} \nonumber\\
&&\qquad {\rm tr}\, \left\{Q(-24E\right.+ 24\chi E \chi +48\chi E +48 E\chi   \nonumber\\
&&\qquad -12\chi_{:a}\chi^{:a}+12\chi_{:a}^{\ a}- 6\chi_{:a}\chi^{:a}\chi +192S^2  \nonumber \\
&&\qquad +96KS +(3+10\chi)K^2+(6-4\chi)K_{ab}K^{ab} )  \nonumber\\
&&\qquad \left. +Q_{;n}(96S+192S^2) +24\chi Q_{;nn}\right\} \label{a3}
\end{eqnarray}
with $m=\dim \mathcal{M}$.

We have to explain the notations. $n$ denotes the inward pointing unit normal vector, while $a,b,c$ correspond to the tangential directions. $h$ is induced metric on the boundary, it is used to contract the tangential indices. The semicolon is used to abbreviate full covariant derivatives, e.g. $Q_{;n}=[\hat \nabla_n,Q]$. The colon is used to denote boundary covariant derivatives. The only difference to full covariant derivatives is that they use the Christoffel symbol corresponding to the boundary metric $h$ when vector indices are differentiated. (The difference is measured by the extrinsic curvature $K_{ab}$). $S$ is the zeroth order term in Robin boundary conditions, see (\ref{Pip}). $\chi$ is the difference of two boundary projectors, $\chi=\Pi_+-\Pi_-$. In our case, $\chi=\ii \varepsilon_\alpha \gamma^5\gamma^n$. Let us remind, that we have assumed that $\mathcal{M}$ is flat, so that all terms containing the Riemann curvature of the bulk metric have been suppressed. 

These formulas are valid for $\tD$ upon the replacement $\mathcal{M}\to \tM$.


\begin{thebibliography}{99}
\bibitem{Niemi:1983rq}
  A.~J.~Niemi and G.~W.~Semenoff,
  Phys.\ Rev.\ Lett.\  {\bf 51}, 2077 (1983).

\bibitem{Redlich:1983kn} 
  A.~N.~Redlich,
  Phys.\ Rev.\ Lett.\  {\bf 52}, 18 (1984).
  doi:10.1103/PhysRevLett.52.18
	A.~N.~Redlich,
  Phys.\ Rev.\ D {\bf 29}, 2366 (1984).
  doi:10.1103/PhysRevD.29.2366
	
\bibitem{AlvarezGaume:1984nf} 
  L.~Alvarez-Gaume, S.~Della Pietra and G.~W.~Moore,
  Annals Phys.\  {\bf 163}, 288 (1985).
  doi:10.1016/0003-4916(85)90383-5

\bibitem{Dunne:1998qy} 
  G.~V.~Dunne,
  Aspects of Chern-Simons theory, in {\it Topological aspects of low dimensional systems},
	Eds.\ A.~Comtet et al, (Springer, Berlin, 1999)
  hep-th/9902115.

\bibitem{Tkachov:2016}
G.~Tkachov,
{\it Topological Insulators}
(Taylor and Francis, Boca Raton, 2016).

\bibitem{Mulligan:2013he} 
  M.~Mulligan and F.~J.~Burnell,
  Phys.\ Rev.\ B {\bf 88}, 085104 (2013)
  doi:10.1103/PhysRevB.88.085104
  [arXiv:1301.4230 [cond-mat.str-el]].

\bibitem{Zirnstein:2013tba} 
  H.~G.~Zirnstein and B.~Rosenow,
  Phys.\ Rev.\ B {\bf 88}, no. 8, 085105 (2013)
  doi:10.1103/PhysRevB.88.085105
  [arXiv:1303.2644 [cond-mat.mes-hall]].

\bibitem{Konig:2014ema} 
  E.~J.~K\"onig, P.~M.~Ostrovsky, I.~V.~Protopopov, I.~V.~Gornyi, I.~S.~Burmistrov and A.~D.~Mirlin,
  Phys.\ Rev.\ B {\bf 90}, no. 16, 165435 (2014)
  doi:10.1103/PhysRevB.90.165435
  [arXiv:1406.5008 [cond-mat.mes-hall]].
	
\bibitem{Kirsten:2001wz} 
  K.~Kirsten,
  {\it Spectral functions in mathematics and physics},
  (Chapman \& Hall/CRC, Boca Raton, FL, 2001).
	
\bibitem{Vassilevich:2003xt} 
  D.~V.~Vassilevich,
  Phys.\ Rept.\  {\bf 388}, 279 (2003)
  doi:10.1016/j.physrep.2003.09.002
  [hep-th/0306138].
	
\bibitem{Gilkey:2004dm} 
  P.~Gilkey,
  {\it Asymptotic formulae in spectral geometry},
	(CRC Press, Boca Raton, 2004). 
	
\bibitem{Chodos:1974je} 
  A.~Chodos, R.~L.~Jaffe, K.~Johnson, C.~B.~Thorn and V.~F.~Weisskopf,
  Phys.\ Rev.\ D {\bf 9}, 3471 (1974).
  doi:10.1103/PhysRevD.9.3471
A.~Chodos, R.~L.~Jaffe, K.~Johnson and C.~B.~Thorn,
  Phys.\ Rev.\ D {\bf 10}, 2599 (1974).
  doi:10.1103/PhysRevD.10.2599
\bibitem{Branson:1992}
T.~P.~Branson and P.~B.~Gilkey,
Diff. Geom. and its Applications {\bf 2}, 249 (1992).

\bibitem{Fursaev:2011zz}
  D.~Fursaev and D.~Vassilevich,
  {\it Operators, Geometry and Quanta : Methods of spectral geometry in quantum field theory}
	(Springer, Dordrecht, 2011)
  doi:10.1007/978-94-007-0205-9  
	
\bibitem{Deser:1997nv} 
  S.~Deser, L.~Griguolo and D.~Seminara,
  Phys.\ Rev.\ Lett.\  {\bf 79}, 1976 (1997)
  doi:10.1103/PhysRevLett.79.1976
  [hep-th/9705052].

\bibitem{Niemi:1986kh} 
A.~J.~Niemi and G.~W.~Semenoff,
  Phys.\ Rev.\ Lett.\  {\bf 54}, 2166 (1985).
  doi:10.1103/PhysRevLett.54.2166;
  A.~J.~Niemi,
  Phys.\ Rev.\ Lett.\  {\bf 57}, 1102 (1986).
  doi:10.1103/PhysRevLett.57.1102

\bibitem{Fosco:2017vxl}
C.~Fosco, G.~L.~Rossini and F.~A.~Schaposnik,
  Phys.\ Rev.\ Lett.\  {\bf 79}, 1980 (1997)
  Erratum: [Phys.\ Rev.\ Lett.\  {\bf 79}, 4296 (1997)]
  doi:10.1103/PhysRevLett.79.1980
  [hep-th/9705124];
  C.~D.~Fosco and F.~A.~Schaposnik,
  arXiv:1702.08229 [hep-th].


\bibitem{Deser:1997gp} 
  S.~Deser, L.~Griguolo and D.~Seminara,
  Phys.\ Rev.\ D {\bf 57}, 7444 (1998)
  doi:10.1103/PhysRevD.57.7444
  [hep-th/9712066].
	
\bibitem{Laughlin:1981jd} 
  R.~B.~Laughlin,
  Phys.\ Rev.\ B {\bf 23}, 5632 (1981).

\bibitem{Halperin:1981ug} 
  B.~I.~Halperin,
  Phys.\ Rev.\ B {\bf 25}, 2185 (1982).
  doi:10.1103/PhysRevB.25.2185


\bibitem{Vafek:2011zh} 
  O.~Vafek,
  Phys.\ Rev.\ B {\bf 84}, 245417 (2011)
  doi:10.1103/PhysRevB.84.245417
  [arXiv:1110.2508 [cond-mat.mes-hall]].

\bibitem{Coste:1989wf} 
  A.~Coste and M.~L\"uscher,
  Nucl.\ Phys.\ B {\bf 323}, 631 (1989).
  doi:10.1016/0550-3213(89)90127-2
	
\bibitem{Kalinichenko:2017nbz}
  I.~S.~Kalinichenko and P.~O.~Kazinski,
  {\it High-temperature expansion of the one-loop effective action induced by scalar and Dirac particles},
  arXiv:1705.01018 [hep-th].
	
\bibitem{mixedBC}
T.~P.~Branson and P.~B.~Gilkey,
  Commun.\ Part.\ Diff.\ Eq.\  {\bf 15}, 245 (1990).
  doi:10.1080/03605309908820686
D.~V.~Vassilevich,
  J.\ Math.\ Phys.\  {\bf 36}, 3174 (1995)
  doi:10.1063/1.531021
  [gr-qc/9404052].
T.~P.~Branson, P.~B.~Gilkey, K.~Kirsten and D.~V.~Vassilevich,
  Nucl.\ Phys.\ B {\bf 563}, 603 (1999)
  doi:10.1016/S0550-3213(99)00590-8
  [hep-th/9906144].
	
\bibitem{Marachevsky:2003zb} 
  V.~N.~Marachevsky and D.~V.~Vassilevich,
  Nucl.\ Phys.\ B {\bf 677}, 535 (2004)
  doi:10.1016/j.nuclphysb.2003.11.009
  [hep-th/0309019].

\end{thebibliography}
\end{document}